\documentclass[prd,twocolumn,letterpaper,superscriptaddress]{revtex4}
\include{epsf}
\usepackage{graphicx}
\usepackage{pslatex}
\usepackage{url}

\begin{document}

\title{Observational Constraints on the Ultra-high Energy Cosmic Neutrino Flux\\
from the Second Flight of the ANITA Experiment}


\author{P.~W.~Gorham$^1$,
P.~Allison$^1$,
B.~M.~Baughman$^2$,
J.~J.~Beatty$^2$,
K.~Belov$^3$, 
D.~Z.~Besson$^4$,
S.~Bevan$^5$,
W.~R.~Binns$^6$,
C.~Chen$^7$,
P.~Chen$^7$,
J.~M.~Clem$^8$,
A.~Connolly$^5$,
M.~Detrixhe$^4$,
D.~De~Marco$^8$,
P.~F.~Dowkontt$^6$,
M.~DuVernois$^1$,
E.~W.~Grashorn$^2$,
B.~Hill$^1$,
S.~Hoover$^3$,
M.~Huang$^7$
M.~H.~Israel$^6$,
A.~Javaid$^8$,
K.~M.~Liewer$^{9}$,
S.~Matsuno$^{1}$,
B.~C.~Mercurio$^2$,
C.~Miki$^{1}$,
M.~Mottram$^5$,
J.~Nam$^7$,
R.~J.~Nichol$^5$,
K.~Palladino$^2$,
A.~Romero-Wolf$^1$,
L.~Ruckman$^1$,
D.~Saltzberg$^3$,
D.~Seckel$^8$,
G.~S.~Varner$^{1}$,
A.~G.~Vieregg$^3$,
Y.~Wang$^7$
}
\vspace{2mm}
\noindent
\affiliation{
$^1$Dept. of Physics and Astronomy, Univ. of Hawaii, Manoa, HI 96822. 
$^2$Dept. of Physics, Ohio State Univ., Columbus, OH 43210. 
$^3$Dept. of Physics and Astronomy, Univ. of California Los Angeles, CA 90095.
$^4$Dept. of Physics and Astronomy, Univ. of Kansas, Lawrence, KS 66045. 
$^5$Dept. of Physics and Astronomy, Univ. College London, London, United Kingdom.
$^6$Dept. of Physics, Washington Univ. in St. Louis, MO 63130. 
$^7$Dept. of Physics, 
National Taiwan Univ., Taipei, Taiwan.
$^8$Dept. of Physics, Univ. of Delaware, Newark, DE 19716. 
$^9$Jet Propulsion Laboratory, Pasadena, CA 91109. 
}

\begin{abstract}
The Antarctic Impulsive Transient Antenna (ANITA) completed its second 
Long Duration Balloon flight in January 2009, with 31 days aloft (28.5 live days) over 
Antarctica. ANITA searches for impulsive coherent radio Cherenkov emission
from 200 to 1200~MHz, arising from the Askaryan charge excess in
ultra-high energy neutrino-induced cascades within Antarctic ice. This flight
included significant improvements over the first flight in payload sensitivity,
efficiency, and flight trajectory. Analysis of in-flight 
calibration pulses from surface and sub-surface locations verifies the expected 
sensitivity. In a blind analysis, we find 2 surviving events on a background,
mostly anthropogenic, of $0.97\pm0.42$ events.  We set the strongest 
limit to date for $10^{18}-10^{21}$~eV cosmic neutrinos,
excluding several current cosmogenic neutrino models. 
\end{abstract}
\pacs{95.55.Vj, 98.70.Sa}
\narrowtext 

\maketitle

\section{Introduction}
The existence of cosmic-ray particles of energies above $10^{19}$~eV, first established
in the early 1960's, has become a problem of the first rank in particle astrophysics.
Models for their production must generate particle energies many orders of
magnitude higher than achievable on Earth, and these models in turn
require extreme source physics that has not yet been formulated in a self-consistent 
manner. Even more problematic, the propagation of such particles 
is limited by strong energy loss due to the Greisen-Zatsepin-Kuzmin 
(GZK) process~\cite{Greisen}. Hadrons are produced via the Delta-photoproduction 
resonance by interactions with cosmic microwave background (CMB) 
photons.
This GZK cutoff in energy
limits the propagation distance of the ultra-high energy cosmic rays (UHECRs) to within 100-200~Mpc 
in the current
epoch, and severely distorts the observed energy spectrum. Astronomy
using charged-particle UHECRs is thus limited to the local universe, suffering
from both the loss of source spectral information, and the difficulty 
in back-tracing UHECRs through intergalactic magnetic fields.

At distances
of several Gpc, corresponding to the star-formation maximum at redshift $z \sim 1$,
the higher energy and density of the CMB photons leads to even greater restrictions
on cosmic-ray propagation, and a more rapid energy loss to Delta photoproduction.
However, information about the source particles does survive in the form of secondary
neutrinos in the decay chain, known as the ultra-high energy (UHE) cosmogenic neutrinos,
first described by Berezinsky \& Zatsepin (BZ)~\cite{BZ}. Their momenta are unaffected by 
magnetic fields, and they propagate without energy loss directly to Earth, retaining 
information about the cosmic distribution of UHECRs and their sources.

The ANITA Long Duration Balloon experiment was designed to search for
cosmogenic neutrinos via electromagnetic cascades initiated by the neutrinos 
in Antarctic ice, the most massive body of accessible, solid, 
radio-transparent dielectric material on Earth. We previously
placed limits on the UHE cosmic neutrino flux from the first flight of ANITA~\cite{ANITA-1}, 
and provided a separate
detailed description~\cite{ANITA-inst} of the ANITA instrument, flight system,
data acquisition, and analysis methods. 
In this
article we detail upgrades and augmentations beyond the
instrument and methodology previously reported.

\section{Experimental Technique}
\begin{figure}[ht!]
 \begin{center}
\includegraphics[width=2.85in]{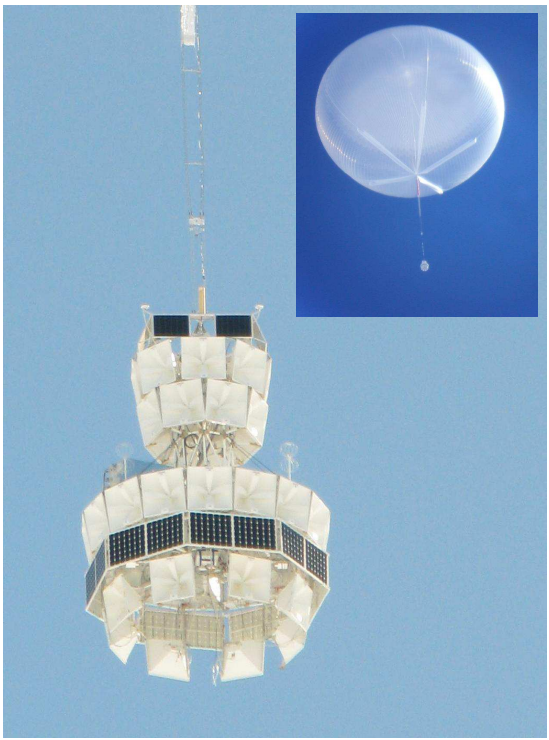}
 \caption{The ANITA-II payload on ascent with the lower eight
horn antennas deployed. The payload
height is $\sim$8 meters, and each antenna face is
0.95~m across. The inset shows the balloon and payload
viewed telescopically 
at float altitude of 35~km.}
 \label{payload08}
 \end{center}
 \end{figure}

The second flight of ANITA (ANITA-II) launched from Williams Field, Antarctica
on December 21, 2008 and landed near Siple Dome after 31 days aloft, resulting in
28.5 live days.
Fig.~\ref{payload08} shows an image
of the payload on ascent after it had deployed to its full flight configuration,
and an inset of the balloon and payload at float altitude.
The mean ice depth in the field-of-view was 1.4~km, 
approximately one attenuation length at sub-GHz radio
frequencies~\cite{icepaper}. 
ANITA-II flew at an altitude of 35-37~km above sea level (33-35~km above 
the ice surface), and was thus able to synoptically view a volume of $\sim1.6$M~$\mathrm{km}^{3}$ 
of ice.
ANITA-II's sensitivity to cosmogenic
neutrinos was improved substantially compared to ANITA-I: 
the front-end system noise temperature was reduced by
40~K, a 20\% improvement in temperature~\cite{ANITA-inst}; 
8 additional quad-ridged horn antennas were added to the previous total of 32;
and the efficiency of the hardware trigger was optimized for impulsive signals.
Also, the instrument was made much more robust to the effects of 
bursts of anthropogenic radio-frequency (RF) interference with the ability to
mask channels from the trigger in the azimuthal sectors of the payload pointing at the noise
source.   Masking occurred on time scales of 
a minute.
This upgrade significantly improved the livetime when in view of strong sources
such as McMurdo and Amundsen-Scott Stations, and trigger thresholds remained at
thermal-noise levels throughout the flight.
The combined effect of all of these modifications led to an increase of about a factor
of four improvement in the expected signal from typical cosmogenic neutrino models as compared
to ANITA-I.

The ANITA-II hardware trigger selects impulsive radio signals 
with broadband frequency content and temporal coherence over nanosecond 
timescales in the vertical polarization.  The broadband nature of triggered signals
was achieved by requiring power in multiple frequency bands, and temporal
coherence was ensured by requiring impulsive power in a full-band channel.
All simulations and laboratory measurements of Askaryan signals~\cite{Ask62, alvarez, Sal01} 
from ice sheet neutrino interactions
show that the RF signal at the payload
is predominantly vertically polarized, due to the
surface Fresnel coefficient and Cherenkov geometry~\cite{ANITA-inst,slac07}.
No strict pulse shape requirements are enforced in the
trigger to allow for variations 
in the shape of any individual neutrino-induced cascade, and the
trigger is thus very inclusive.
The trigger threshold rides at the ambient
thermal noise level to maintain an approximately constant
trigger rate of $\sim$10~Hz, which is dominated~(98.5\%) by incoherent
thermal noise fluctuations. These thermal-noise triggers 
have well-modeled statistical probabilities
for producing random impulses, and are highly suppressed in analysis with
requirements of spatial and temporal coherence. Most other triggers
are from anthropogenic sources. Such signals can mimic neutrino-like
impulses, and may arise from high-voltage
discharges in electrical equipment or from metallic structures 
charged by blowing snow or related effects. To remove such
backgrounds, we identify active and prior human 
activity in Antarctica
and optimize pointing resolution to reliably associate
anthropogenic signals with known sources.
We know of no expected particle-physics backgrounds.

Extensive calibration and validation 
of the system response and trigger efficiency 
for a range of impulse signal-to-noise ratios (SNRs) was
done both prior to flight and with ground-to-payload
impulse generating antennas during flight. These in-flight pulser
systems were located at the launch site and at
a remote field station at Taylor Dome on the edge of the Antarctic plateau.
In both cases, impulse generators were fed through an antenna
immersed in the ice as deep as 90~m~\cite{ANITA-inst}. In-flight measurements of the
impulses from Taylor Dome
provided validation that
refraction effects on signal propagation through the ice surface 
do not significantly affect the coherence of the received signal, 
to distances of 400~km. Measurements of trigger 
efficiency in-flight were consistent with expectations 
from the ground calibrations, considering the narrower frequency content
of impulses from Taylor Dome.
Ground-to-payload signals provided an equally critical function in detector alignment 
and in determining
the precision of directional reconstruction.
ANITA's antenna signals are combined via pulse-phase interferometric
methods~\cite{ANITA-inst}, resulting in a radio map (``interferometric image'') for each polarization
of the intensity as a function
of payload elevation and azimuth. The
largest peak in either map determines the direction of the signal source. The
payload coordinate frame is tied to the geodetic frame via
onboard GPS and sun-sensor measurements to a precision of about
0.1$^{\circ}$. The pointing resolution in our analysis is $0.2^{\circ}-0.4^{\circ}$ in
elevation and $0.5^{\circ}-1.1^{\circ}$ in azimuth, depending on the SNR of the event.

\section{Data Analysis}
Our signal region comprises events that do not come from known sources of
human activity (``camps'') and have no geospatial partner event within
distance and angular separation criteria.
We blinded the signal region in two ways simultaneously.  First, we inserted an undisclosed number
(12) of neutrino-like
calibration events at random, concealed times throughout the flight.  Second, 
we blinded ourselves to events in the signal region {\it i.e.} events that have no spatial partner.
The analysis efficiency and background are estimated
before opening this hidden signal box.
Prior to unblinding, the complete set of in-flight calibration pulser data is used to optimize the
pointing reconstruction of the analysis.  In this section, we describe
the sequence of analysis cuts that we applied to the sample until the final
step, opening the blind signal box.
Table~\ref{event-table} shows the the total event 
sample, the sequential number of surviving events as each analysis cut is applied,
and the efficiency of each cut.
The number of {\bf Hardware-Triggered Events} is the total number of events recorded. 
{\bf Quality Events} exclude on- and off-payload calibration pulser events, unbiased
auto-trigger events, and a small fraction ($\sim$1\%) of events with data corruption.
\begin{table}[htb2!]
\caption{\label{event-table}Event totals vs. analysis cuts and
estimated signal efficiencies for ESS spectral shape~\cite{Engel01}.
}
\begin{center}
\begin{tabular}{lccc}
\hline
Cut requirement   &\multicolumn{2}{c}{Passed}  & Efficiency \\
                 &~~~~~Vpol~~~~~&~~~~~Hpol~~~~~& \\
\hline
~~~~Hardware-Triggered Events & \multicolumn{2}{c}{$\sim 26.7$M}& -\\
(1) Quality Events & \multicolumn{2}{c}{$\sim 21.2$M}& 1.00\\
(2) Reconstructed Events & \multicolumn{2}{c}{320,722}& 0.96\\
(3) Not Traverses and Aircraft & \multicolumn{2}{c}{314,358}& 1.00\\
(4) In Clusters $<$100 Events & \multicolumn{2}{c}{444}& -\\
(5) Isolated Singles & 7 & 4 & 0.64\\
(6) Not Misreconstructions & 5 & 3 & 1.00\\ %
(7) Not of Payload Origin & 2 & 3 & 1.00\\
\hline
Total Efficiency &  &  & 0.61\\
\hline
\end{tabular}
\end{center}
\end{table}
\begin{figure}[ht!]
  \begin{center}
    \includegraphics[width=3.3in]{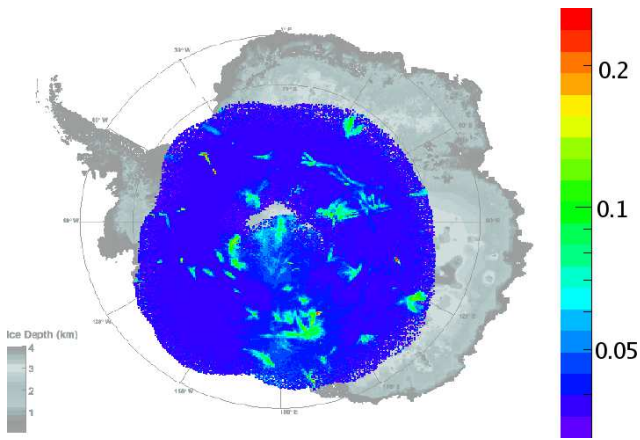}
    \caption{The vertical-polarization reconstruction position of all 21.2~M Quality Events,
      with the color scale representing the average peak value of the 
      interferometric image for events which fall in
      that bin.
      The dark blue region is consistent with pure thermal noise.  The green, yellow, 
      and red
      regions are increasingly non-thermal and are consistent with anthropogenic noise.}
    \label{fig:rfmap}
  \end{center}
\end{figure}

The great majority of remaining events are random thermal-noise-coincidence events 
in which individual waveforms are impulsive, but the ensemble of signals are incoherent 
and uncorrelated between antennas. 
We create an interferometric image in each polarization by cross-correlating waveforms
from neighboring antennas and summing the total normalized cross-correlation
value for each elevation and azimuth.  We construct a ``coherently summed'' 
waveform given the direction of the largest peak in either map using the antennas that
are closest to that peak.  Fig.~\ref{fig:rfmap} shows the vertical-polarization
reconstruction position of all 21.2~M Quality Events.  The color 
of each bin represents the average value of the peak of the interferometric image
for events which fall in that bin.  A value of $<0.05$ corresponds to the 
thermal-noise floor. The clear separation of thermal
and anthropogenic events motivates our accounting for each background separately.  Nearly
all of the non-thermal excess will be removed in analysis because of its association with
known bases.

Thermal noise power originates from the ice and the electronics.  Because thermal noise
does not correlate between antennas, the peak in the interferometric image is equally likely to
be in any direction.
We establish thermal noise cuts based on a sample of 
$2.1$~M above-horizon (non-neutrino) events from a quiet period during the flight, when ANITA
was not in view of McMurdo Station.  
To establish {\bf Reconstructed Events},
we reject thermal noise by cutting on the 
peak value of the normalized cross-correlation in the interferometric image, the peak of the 
envelope of the coherently summed waveform, and its fraction of linear polarization. 
We also require that events reconstruct 
to locations where there is ice on the ground, down to $35^{\circ}$ below horizontal.  
Using a modest extrapolation verified by simulation,
we reject thermal noise events at a 
level of $\leq 2.5\times10^{-8}$, corresponding to a background of $0.50 \pm 0.23$~events out 
of the 21.2~M Quality Events. 
The uncertainty represents a variation of fit parameters used for the extrapolation.
For more detail, see Reference~\cite{ViereggThesis}.
Events which
pass this set of cuts have been filtered for the presence of continuous-wave (CW) interference --
typically narrow frequency-band signals from radio transmitters -- and are
largely impulsive in nature.

Remaining events are mostly anthropogenic. 
To cut anthropogenic noise at a level of $\leq 3\times10^{-6}$, 
we require that any neutrino candidate be a single, geospatially-isolated event.
It is extremely
unlikely for two neutrino events to occur in close proximity to one another; 
correlated event locations are likely to indicate human activity.
We first remove all events associated in both time and space
with known flights and over-land traverses, so the remaining are 
{\bf Not Traverses and Aircraft}.  We cluster all remaining 
events with known active and inactive camps, other events that also pass all previous cuts,
and locations where 
low-level RF power has been detected by statistical correlation 
of weak signals, shown in Fig.~\ref{fig:rfmap}. 
Events are clustered with a base or another event if
the distance between the reconstructed location(s) is $<$40~km or if
the angular separation between the locations is $<5.5$ times the pointing resolution for
the event(s).  Table~\ref{mult-table} shows the number of clusters of reconstructed events vs. cluster 
multiplicity for {\bf Clusters $<$100 Events} to give a sense of the distribution of anthropogenic noise,
and is used to estimate the anthropogenic background contribution.
Remaining unclustered events are {\bf Isolated Singles}.
The isolation requirement lowers acceptance
since each event removes a
region of the ice sheet around it from the available neutrino target volume.
We used seven largely independent methods to estimate that the anthropogenic background remaining
after our clustering cuts is $0.65\pm 0.39$ vertically-polarized (Vpol) events and $0.25\pm0.19$ 
horizontally polarized (Hpol) events~\cite{ViereggThesis}.  (For example, in Table~\ref{mult-table}, 
there are 7 single events from known bases, 3 non-single clusters not from known bases, 
and 17 non-single clusters from known bases, yielding $3 \times 7 / 17 = 1.2$.  After
the polarization cut, this estimate becomes 0.91 in Vpol and 0.28 in Hpol.  Combining this estimate
with six other estimates that use a similar technique gives the quoted value and uncertainty.)
 \begin{table}[htb3!]
 \caption{\label{mult-table}Event cluster multiplicities for all
 clusters with fewer than 100 reconstructed events.  All the cuts in
 Table~\ref{event-table} have been applied except for the Isolated Singles cut.  The
 5 events in the signal region correspond to the 2+3 events in the last entry in 
 Table~\ref{event-table}.
 }
 \begin{center}
 \begin{tabular}{lcc}
 \hline
 Cluster Multiplicity~~~~~~~& \multicolumn{2}{c}{Number of Clusters} \\
 &Camp~~~~~&Not Camp~\\ \hline
 10-100 & 8 & 1 \\
 5-9 & 7 & 1 \\
 4 & 1 & 0 \\
 3 & 0 & 0 \\
 2 & 1 & 1 \\
 1 & 7 & 5~(Signal Region)\\ \hline
 \end{tabular}
 \end{center}
 \end{table}

At this stage, the polarization angle of the event is calculated
using the Stokes parameters, and the event is assigned to be 
Hpol ($<40^{\circ}$), Vpol ($>50^{\circ}$), or sideband ($40^{\circ}-50^{\circ}$). 
There were no events remaining in
the sideband.  Events that are {\bf Not Misreconstructions} 
have a low probability of misreconstruction;
we remove any event that clearly peaked at 
a sidelobe of the pattern in the interferometric image. 
This cut was tested using events from the ground-to-payload calibration pulsers
and known camps.
The requirement that events must be {\bf Not of Payload Origin} removes
events associated with local interference originating on the
payload but missed in the quality event stage; these are easily identified.

Upon opening the blind box, we first examine what happened to the 
12 undisclosed, inserted neutrino-like events.  Of the 12 inserted 
events, 11 were unique events with one duplicate event.  Of the 
11 unique events, 8 were Reconstructed Events, consistent with
the calculated reconstruction analysis efficiency for such low-SNR events.  Expected
neutrino events have a wide range of SNR, but always at least as large as the inserted 
events.

\section{Results}
\begin{figure}[ht!]
\begin{center}
\includegraphics[width=3.3in]{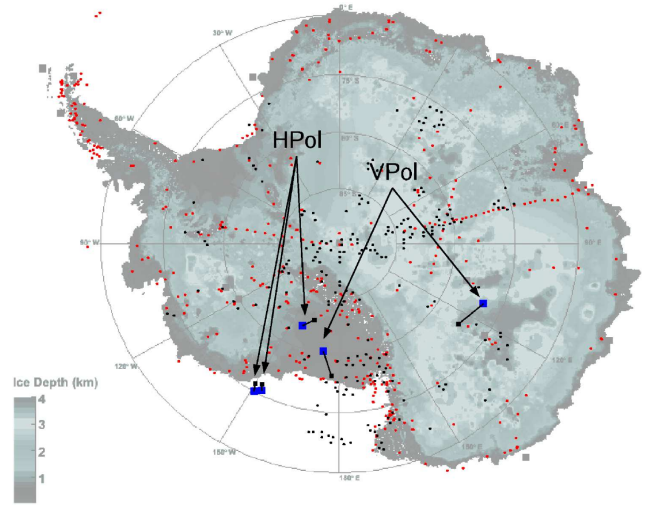}
\caption{
Events remaining after unblinding.  The Vpol neutrino
channel contains two surviving events. 
Three candidate UHECR events remain in the Hpol channel. Ice depths are from BEDMAP~\cite{BEDMAP}.}
\label{eventmap}
\end{center}
\end{figure}

After all cuts are applied, two events remain in the Vpol channel, and three in the Hpol
channel.
After clustering cuts, the thermal noise background 
reduces to $0.32\pm0.15$ in each channel.
The total background is $0.97\pm0.42$ 
events in the Vpol channel, and $0.67\pm0.24$ events in the Hpol channel.
Thermal noise backgrounds are likely reducible in future analysis.
Fig.~\ref{eventmap} shows the reconstructed locations of the remaining events (large blue squares)
on the
Antarctic continent and the payload position at the time of detection
(small black square connected to the blue square), along with camps (red points), 
and locations of low-level RF noise (black points).
\begin{figure}[ht!]
\begin{center}
\includegraphics[width=3.3in]{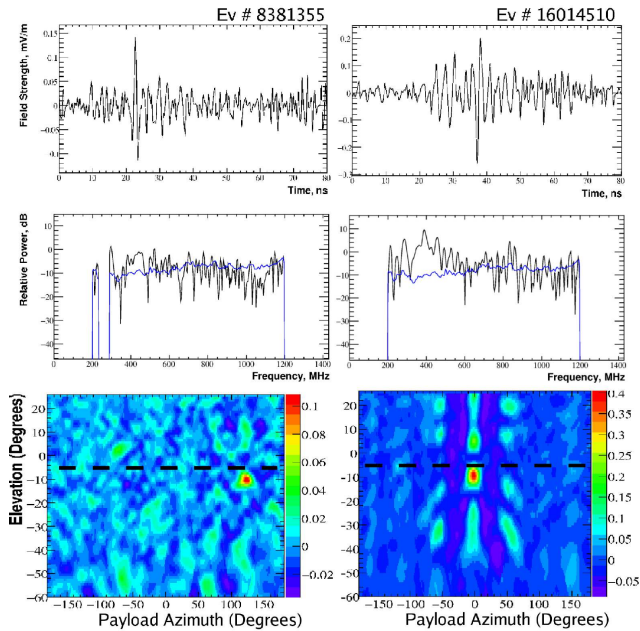}
\caption{ Top: Waveforms of incident field strength for the two surviving Vpol events.  
Event 8381355 is shown filtered between 235-287~MHz to remove weak CW noise.  
Middle: Corresponding frequency power spectra with an average thermal-noise spectrum shown in blue for
reference.
Bottom: Corresponding interferometric images showing the pulse direction. The dashed line is the horizon.}
\label{events}
\end{center}
\end{figure}

All three Hpol events show characteristics
which identify them as geo-synchrotron 
radio emission from UHECR air showers, reflecting from the ice surface (including sea ice), 
as described in our ANITA-I results~\cite{ANITA1nature}.
While ANITA-I saw 16 such events, the much smaller number
of Hpol events seen in ANITA-II is due to the change of the
trigger to favor Vpol events to maximize neutrino sensitivity. 

The two remaining Vpol events are of unknown origin. 
In Fig.~\ref{events} we show some of the characteristics
of these events, including the waveforms, frequency spectra, and interferometric
images.  Event 8381355
has been filtered using the adaptive filter developed for the analysis and 
is highly impulsive, with a nearly flat radio spectrum. Event 16014510
shows a central impulse with some additional distributed signal within 10-15~ns of the peak,
and a frequency spectrum peaking near 400~MHz.
The event still passes all cuts
if the 400~MHz region is filtered by hand.
The reconstructed directions are robust, supporting
identification as isolated events. The waveforms and frequency spectra are
within the range of simulated neutrino events.
Both events are consistent in their locations and
amplitudes with distributions of Monte-Carlo-generated neutrinos.
We lack adequate statistics to identify these two events as a unique
non-anthropogenic population.  

We proceed to set a limit including systematic errors~\cite{fc}.
The largest systematic error
is on the acceptance, which is calculated using two independent Monte Carlo simulations~\cite{ANITA-inst}.
The two simulations typically differ by 20\%, which we take as a systematic error.
The uncertainty on analysis efficiency includes two effects: statistical uncertainty on the 
efficiency calculation using calibration pulses, and a 
systematic error from the comparison of the efficiency on simulated neutrino events and calibration events.  
This uncertainty on analysis efficiency is 3\%. The uncertainty on the background is discussed above.  
The inclusion of systematic errors only worsens the limit by
about 10\% because the procedure given in Ref.~\cite{fc} accounts for both signs of systematic fluctuations.  

Two other systematic sources are theoretical in nature.  Changes in cross section, $\sigma$, can affect the limit
in two ways: increasing the neutrino-nucleon cross section increases the interaction rate, but lowers
the solid angle due to Earth shielding (and {\it vice-versa}).  The net result is an event rate which
scales as $\sigma^{0.45}$.  For a choice of cross section different
than used here~\cite{grqs}, our limit can be adjusted accordingly.  Similarly,
the event rate depends linearly on the Askaryan electric field~\cite{alvarez}.  Including reasonable
variations on these parameters would affect the limit by $\sim10$\%.

\section{Discussion}
\begin{figure}[ht!]
\begin{center}
\includegraphics[width=3.2in]{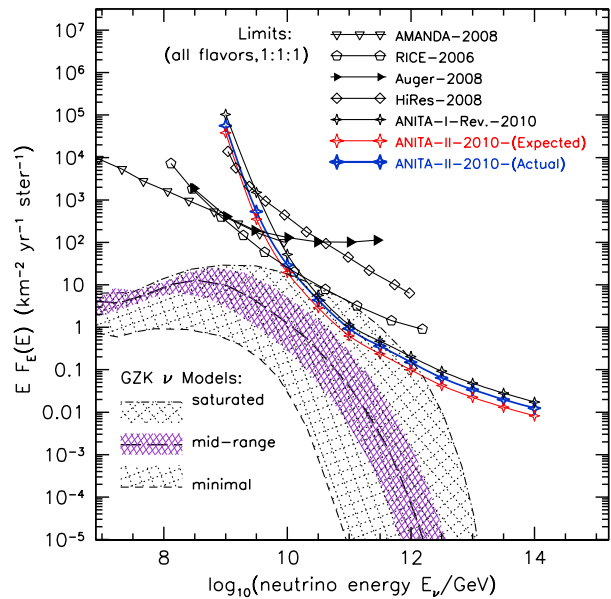} \\
\caption{ANITA-II limit for 28.5 days livetime. The red curve
is the expected limit before unblinding, 
based on seeing a number of candidates equal to the background estimate. The blue
curve is the actual limit, based on the two surviving candidates. Other
limits are from AMANDA~\cite{AMANDA08}, RICE~\cite{RICE06},
Auger~\cite{Auger07}, HiRes~\cite{Hires08}, and a revised limit from ANITA-I~\cite{hooverThesis}. 
The BZ (GZK) neutrino model range is determined by a variety of 
models~\cite{PJ96,Engel01,Kal02,Kal02a,Aramo05,Ave05,Barger06}.}
\label{lim10}
\end{center}
\end{figure}

Our model-independent~\cite{Anch02, FORTE04} 90\%~CL limit 
on neutrino fluxes is based on the 28.5 day livetime,
energy-dependent analysis efficiency (68\%-42\% from $10^{18}$-$10^{23}$~eV), 
the average acceptance from the two 
independent simulations,~\cite{ANITA-inst}\footnote{The acceptance used 
from $10^{18}$-$10^{23}$~eV in
half-decade energy steps
is: $4.3\times10^{-4}$, $5.0\times10^{-2}$, 0.92, 6.6, 36., 
108., 259., 602., 1140, 1950, and 3110~km$^2$-sr}
and $0.97\pm0.42$ expected background 
events including the
systematic effects described above.
Relative to the revised ANITA-I limit~\cite{hooverThesis,ANITA-1} shown in Fig.~\ref{lim10},
the {\it expected} limit from this data, in the absence of signal, is a factor of four more sensitive.
We set the {\it actual} limit, shown
in Fig.~\ref{lim10}, using our 2 observed candidates.  Because ANITA-II saw more than
the expected background, 
the {\it actual} limit is only a factor of two better than ANITA-I even 
though the {\it a priori} sensitivity is four times higher for ANITA-II. 
The ANITA-II limit supercedes the ANITA-I limit and would
not significantly be improved by combining the results.

Table~\ref{table2} gives integrated event totals for a
range of cosmogenic neutrino models with widely varying assumptions.
We also include for reference the expected number of events
for a pure power-law neutrino spectrum that matches
the Waxman-Bahcall flux bounds for both evolved and standard
UHECR sources~\cite{WB}. ANITA-II's constraint on cosmogenic neutrino models strongly
excludes models with maximally energetic UHECR source spectra
which saturate other available bounds~\cite{Yosh97,Kal02,Aramo05}. 
These models generally assume very flat source energy spectra
which may extend up to $10^{23}$~eV; our results are incompatible
with a combination of both of these features.
ANITA-II is now probing several models with
strong source evolution spectra that
are plausible within 
current GZK source expectations~\cite{Kal02,Aramo05,VB05,Barger06,Yuksel07},
some at $>90$\% confidence level.
The ANITA-II 90\% CL integral flux limit on a pure $E^{-2}$ spectrum
for $10^{18}$~eV $\leq E_{\nu} \leq 10^{23.5}$~eV is 
$E_{\nu}^2 F_{\nu} \leq 2 \times 10^{-7}$~GeV~cm$^{-2}$~s$^{-1}$~sr$^{-1}$. 
These differential and integral limits, as well as the individual model
limits above, are the strongest constraints to date on the cosmogenic
UHE neutrino flux.

\section{Acknowledgments}
We thank the National Aeronautics and Space
Administration, the National Science Foundation Office of Polar Programs, 
the Department of Energy Office of Science HEP Division, the UK Science
and Technology Facilities Council, 
the National Science Council in Taiwan ROC,
and especially the staff of 
the Columbia Scientific Balloon Facility.\vspace*{1.5in}
\begin{table}[hbt!]
\caption{Expected numbers of events $N_{\nu}$ from several cosmogenic neutrino models, and 
confidence level for exclusion by ANITA-II observations when appropriate.
\label{table2}}
\vspace{3mm}
 \begin{footnotesize}
  \begin{tabular}{lcr}
\hline
{ {\bf Model \& references}}   &  predicted {$N_{\nu}$}~~      &  ~~{\bf CL,\%} \\ \hline
{\it Baseline models:} &  &  \\
~~~~~~Various~\cite{PJ96,Engel01,Kal02,Stan06,Barger06} & 0.3-1.1 & ...\\
{\it Strong source evolution models:}&  &  \\
~~~~~~Aramo {\it et al.} 2005~\cite{Aramo05} & 2.6 & 78 \\
~~~~~~Berezinsky 2005~\cite{VB05} & 5.4 &   97 \\
~~~~~~Kalashev {\it et al.} 2002~\cite{Kal02} &  5.9 & 98 \\
~~~~~~Barger, Huber, \& Marfatia 2006~\cite{Barger06} & 3.6 & 89 \\
~~~~~~Yuksel \& Kistler 2007~\cite{Yuksel07} & 1.8 & ... \\
{\it Models that saturate all bounds}: & & \\
~~~~~~Yoshida {\it et al.} 1997~\cite{Yosh97} & 32 & $>99.999$  \\
~~~~~~Kalashev {\it et al.} 2002~\cite{Kal02} &  20 & $>99.999$ \\
~~~~~~Aramo {\it et al.} 2005~\cite{Aramo05} & 17 & 99.999 \\
{\it Waxman-Bahcall fluxes}: & & \\
~~~~~~Waxman, Bahcall 1999, evolved sources~\cite{WB}~~ & 1.5 & ...  \\
~~~~~~Waxman, Bahcall 1999, standard~\cite{WB} & 0.52 & ...   \\ \hline
  \end{tabular}
 \end{footnotesize}
\end{table}

\end{document}